\documentclass[aps,prb,preprint,preprintnumbers,amsmath,amssymb,groupedaddress]{revtex4}
\usepackage{amsfonts,amsmath,amssymb}
\usepackage{graphicx}

\begin{document}

\newlength{\figwidth}
\setlength{\figwidth}{0.8\textwidth}

\preprint{JCP/Sept2004}

\title{Close-coupling calculations of low-energy
inelastic and elastic processes in $^4$He collisions with H$_2$: A
comparative study of two potential energy surfaces}

\author{Teck-Ghee~Lee}
\affiliation{Department of Physics and Astronomy, University of
Kentucky, Lexington, KY 40506 \\ and Physics Division, Oak Ridge
National Laboratory, Oak Ridge, TN 37831}

\author{C.~Rochow, R.~Martin, T.~K.~Clark, and R.~C.~Forrey}
 \affiliation{Department of Physics, Penn State University, Berks-Lehigh Valley College, Reading, PA 19610}

\author{N.~Balakrishnan}
 \affiliation{Department of Chemistry, University of Nevada--Las Vegas, Las Vegas, Nevada 89154 }

\author{P.~C.~Stancil}
 \affiliation{Department of Physics and Astronomy and Center for Simulational Physics,
 University of Georgia, Athens, GA 30602}

\author{D.~R.~Schultz}
 \affiliation{Physics Division, Oak Ridge National Laboratory, Oak Ridge, TN 37831}

\author{A.~Dalgarno}
 \affiliation{ITAMP, Harvard-Smithsonian Center for Astrophysics, 60 Garden Street, Cambridge, MA 02138}

\author{Gary J.~Ferland}
 \affiliation{Department of Physics and Astronomy, University of Kentucky, Lexington, KY40506}

\date{\today}

\begin{abstract}
The two most recently published potential energy surfaces (PESs)
for the HeH$_2$ complex, the so-called MR (Muchnick and Russek)
and BMP (Boothroyd, Martin, and Peterson) surfaces, are
quantitatively evaluated and compared through the investigation of
atom-diatom collision processes. The BMP surface is expected to be
an improvement, approaching chemical accuracy, over all
conformations of the PES compared to that of the MR surface. We
found significant differences in inelastic rovibrational cross
sections computed on the two surfaces for processes dominated by
large changes in target rotational angular momentum. In
particular, the H$_2$($\nu=1, j=0$) total quenching cross section
computed using the BMP potential was found to be a factor of
$~$1000 larger than that obtained with the MR surface. A lesser
discrepancy persists over a large range of energies from the
ultracold to thermal and occurs for other low-lying initial
rovibrational levels. The MR surface was used in previous
calculations of the H$_2$($\nu=1, j=0$) quenching rate coefficient
and gave results in close agreement with the experimental data of
Audibert {\it et al.} which were obtained for temperatures between
50 and 300 K. Examination of the rovibronic coupling matrix
elements, that are obtained following a Legendre expansion of the
PES, suggests that the magnitude of the anisotropy of the BMP
potential is too large in the interaction region. However, cross
sections for elastic and pure rotational processes obtained from
the two PESs differ typically by less than a factor of two. The
small differences may be ascribed to the long-range and anharmonic
components of the PESs. Exceptions occur for ($\nu=10, j=0$) and
($\nu=11, j=1$) where significant enhancements have been found for
the low-energy quenching and elastic cross sections due to
zero-energy resonances in the BMP PES which are not present in the
MR potential.

\end{abstract}

\maketitle


\section{INTRODUCTION}\label{intro}

The study of the He--H$_2$ interaction has a long history and can
be traced back to at least the construction of a simple analytic
potential energy surface (PES) by Roberts \cite{Roberts} in 1963.
Since the beginning of the 70s, scattering between atomic He and
the H$_2$ molecule has been investigated experimentally
\cite{Dove, Audibert, Lukasik}. The laser Raman scattering
technique pioneered by Audibert {\it et al.} \cite{Audibert} has
given accurate and detailed results. These experiments, which
covered the sub-thermal regime, were accurate enough to reproduce
the small structures in the measured rate constants (see
Ref.\cite{Grosjean}) and have prompted considerable theoretical
interest \cite{Alexander,Miller,Billing,Lin,Grosjean}.
Calculations have been performed using {\it ab initio} and
approximate quantal formalisms, classical and semiclassical
methods, and various intermolecular potentials.

Collisions involving hydrogen molecules and helium atoms are of
great interest for three main reasons. First, this collision
system is a prototype for chemical dynamics studies and can be
used as a testing ground for the scattering theory of non-reactive
atom-diatom collisions involving a weak interaction potential.
Second, rotational and vibrational transitions in H$_2$ induced by
collisions with He are of practical importance in models of
astrophysical environments where the physical conditions may not
be accessible to terrestrial experiments. Examples include low
densities characteristic of giant molecular clouds in the
interstellar medium where star formation occurs. Heating of the
interstellar cloud by strong shock waves induces rotational and
vibrational excitation of the H$_2$ molecules leading to
collision-induced dissociation to two free H atoms. Third, with
recent experimental advances in the trapping of molecules
\cite{Doyle,Gould,Pillet,Knize}, collisional studies of the
HeH$_2$ system have given new insight into the behavior of
atom-diatom collisions at ultracold temperatures including
investigations of Feshbach resonances, predissociation in van der
Waals complexes, determination of complex scattering lengths,
testing of effective range theory and Wigner threshold laws, and
quasiresonant vibration-rotation energy transfer
\cite{Bala97,Bala98,Forrey98,Forrey99,Forrey01}.

Due to the continual advancement of computer technology and
computational methodologies, theoretical studies
\cite{Flower98,Bala99} of inelastic collisions between He and
H$_2$ have been performed with increasingly larger basis sets and
have yielded a deeper understanding of the collision mechanism as
well as providing benchmark data for astrophysics and chemistry.
The accuracy of quantum scattering calculations is dictated by the
flexibility and reliability of the interaction potential energy
surface of the projectile atom and target diatom. Especially in
the case of atom-diatom collisions in the ultracold regime, where
the interaction time between the projectile atom and molecular
target is considerably longer than the rotational/vibrational
period of the molecule, the incoming atom is a sensitive probe of
the PES of the complex. Consequently, the target molecule has time
to adjust to the field of the slow moving atom and the scattering
cross sections are sensitive to the anisotropy of the PES.

Muchnick and Russek \cite{MR} constructed a PES for HeH$_2$ which
incorporated the {\it ab initio} potential energy calculations of
Meyer {\it et al.} \cite{Meyer} and Russek and Garcia \cite{RG}. A
crucial feature of the MR PES is the inclusion of the H$_2$
vibrational coordinate, $r$, to describe the motion of the nuclei
in the molecule. The PES was constructed to behave in a physically
realistic fashion in the non-equilibrium regions but the fit was
not constrained by {\it ab initio} data.

The PES calculated by Muchnick and Russek was employed by Flower
{\it et al.} \cite{Flower98} and later by Balakrishnan {\it et
al.} \cite{Bala98,Bala99} and Forrey {\it et al}.
\cite{Forrey98,Forrey99,Forrey01} to obtain the cross sections and
the corresponding rate coefficients for rovibrational transitions
in ortho- and para-H$_2$ induced by collisions with He. Flower
{\it et al.} applied a quantal coupled-channel method that used a
harmonic oscillator approximation for the H$_2$ wave functions and
their presented results for vibrational states $\nu$ = 0, 1 and 2.
Comparison was made with previous calculations and with
measurements at both low and high temperature and the agreement
was found to be good. Balakrishnan {\it et al.} \cite{Bala98}
performed similar calculations with numerically determined H$_2$
wave functions using Hermite basis sets and the H$_2$ potential of
Schwenke \cite{Schwenke}. Their predicted quenching rate
coefficients with the MR PES gave good agreement with the
experimental results of Audibert {\it et al.} \cite{Audibert} for
vibrational relaxation of H$_2$($\nu=1, j=0$) by He impact at
temperatures between 50 and 300 K. Subsequently, the same PES has
been employed to investigate rotational and vibrational excitation
transitions \cite{Bala99,Forrey98,Forrey99,Forrey01}.

The most recent analytic HeH$_2$ PES was constructed by Boothroyd,
Martin and Peterson (BMP) \cite{BMP} from more than 25,000 {\it ab
initio} data points. The BMP surface not only accurately
represents the van der Waals potential well, but also fits the
interaction region with chemical accuracy, giving an order of
magnitude improvement in RMS errors compared to the MR PES. While
such an improvement is critical for chemical dynamics studies, the
BMP potential was also constrained by the {\it ab initio} data to
accurately describe large H$_2$-molecule sizes and for short He
impact distances. These enhancements allow studies of highly
excited H$_2$ and collision-induced dissociation.

In this paper we present a comparative study of the MR and BMP
potential surfaces for collisions of vibrationally and
rotationally excited H$_2$ by He impact. The scattering cross
sections and their corresponding rate coefficients are calculated
using the non-reactive quantum mechanical close-coupling method.
In section \ref{calc}, we outline close-coupling theory and give a
brief description of the PESs. We present our results and
discussion in section \ref{results} and a summary and conclusions
in section \ref{summary}. Atomic units are used throughout, unless
otherwise noted: i.e., $e=m_e=a_o=1$~a.u., while 1 hartree =
27.2116 eV = 627.51 kcal/mol.

\section{THEORETICAL METHODS}\label{calc}
\subsection{The close-coupling approach}

Calculations of rovibrational transition cross sections and
thermally averaged rate coefficients provide stringent tests of
the potential energy surfaces of the HeH$_2$ molecule. To compute
these cross sections and rate coefficients, we use a quantum
mechanical close-coupling method that has been described in detail
elsewhere \cite{Child}. Here we provide a brief overview of the
essential elements of the approach. The time-independent
Schr\"{o}dinger equation for the He+H$_2$ collision system in the
center of mass frame is given by
\begin{equation}
\left(T_r+T_R+v_{H_2}(r)+
  V_{I}(r,R,\theta)-E\right)
  \Psi^{JM}(\vec{R},\vec{r})=0,
  \label{FTISE}
\end{equation}
with $T_r$ = $-\frac{1}{2m}\nabla^{2}_{r}$ and $T_R$ =
$-\frac{1}{2\mu}\nabla^{2}_{R}$ where {\it m} is the reduced mass
of the H$_2$ molecule and $\mu$ is the reduced mass of the
He--H$_2$ complex. The internuclear distance between the two H
atoms is denoted by $r$, $R$ is the distance between the He atom
and the center of mass of H$_2$, and $\theta$ is the angle between
$\vec{r}$ and $\vec{R}$. The term $v_{H_2}$($r$) is the isolated
H$_2$ potential and $V_I$($r,R,\theta$) is the He--H$_2$
interaction energy. To solve eqn.(\ref{FTISE}), we expand the
total wave function $\Psi^{JM}(\vec{R},\vec{r})$ in the form
\begin{equation}
\Psi^{JM}(\vec{R},\vec{r})=\frac{1}{R}\sum_n
C_n(R)\phi_n(\hat{R},\vec{r}),
\label{totalwf}
\end{equation}
where the channel function [$n\equiv(\nu jl;JM)$] is given by
\begin{equation}
\phi_n(\hat{R},\vec{r})=\frac{1}{r}\chi_{\nu j}(r)\sum_{m_j,m_l}
 (j, l, J | m_j, m_l, M)Y^{j}_{m_j}(\hat{r})Y^{l}_{m_l}(\hat{R}).
  \label{chanwf}
\end{equation}
The vibrational and rotational quantum numbers are respectively
denoted by $\nu$ and $j$, and $l$ is the orbital angular momentum
of He with respect to H$_2$, $J$ is the total angular momentum
quantum number (i.e., $\vec{J} = \vec{l}+\vec{j}$), $M$ is the
projection of $J$ onto the space-fixed z--axis, and $(j, l, J|
m_j,m_l, M)$ denotes a Clebsch-Gordon coefficient.  The
corresponding eigenvalues $\epsilon_{\nu j}$ (rovibrational
binding energies) are obtained by solving the radial ($r$) nuclear
Schr\"{o}dinger equation for the diatom, H$_2$,
\begin{equation}
\left(-\frac{1}{2m}\frac{d^2}{dr^2}+
 \frac{j(j+1)}{2mr^2}+v_{H_2}(r)\right)
  \chi_{\nu j}(r)=\epsilon_{\nu j}\chi_{\nu j}(r)
  \label{diatomSE}
\end{equation}
by expanding $\chi_{\nu j}(r)$ in terms of a Hermite polynomial
basis with the H$_2$ potential $v_{H_2}$(r), taken from Schwenke
\cite{Schwenke}. Substituting
eqns.(\ref{totalwf})-(\ref{diatomSE}) into eqn.(\ref{FTISE}), we
arrive at a system of close-coupling equations
\begin{equation}
\left( \frac{d^2}{dR^2}-\frac{l_i(l_i+1)}{R^2}+2\mu
E_i\right)C_i(R) = 2\mu\sum_n
C_n(R)\langle\phi_i|V_I|\phi_n\rangle, \label{cceqn}
\end{equation}
where $E_i =E_{\nu j}$ is the initial kinetic energy and $l_i$ is
the orbital angular momentum in the $i$-th channel. To solve the
coupled radial equations (\ref{cceqn}), we used the hybrid
modified log-derivative-Airy propagator \cite{Manolopoulos} in the
general purpose scattering code MOLSCAT \cite{Molscat}. The
log-derivative matrix \cite{Manolopoulos} is propagated to large
intermolecular separations where the numerical results are matched
to the known asymptotic solutions to extract the physical
scattering matrix. This procedure is carried out for each partial
wave until a converged cross section is reached. We have checked
that the results are converged with respect to the number of
partial waves as well as the matching radius for all channels
included in the calculations.

We also adopt here the total angular momentum representation
introduced by Arthurs and Dalgarno \cite{Dalgarno} in which the
cross section for transitions from an initial $\nu j$
vibrational-rotational level to the final $\nu'j'$ level is given
by
\begin{multline}
\sigma_{\nu j \rightarrow \nu' j'}(E_{\nu j})=
 \frac{\pi}{2\mu E_{\nu j}(2j+1)}
  \sum_{J=0}^{\infty}(2J+1) \\
   \times\sum_{l=|J-j|}^{|J+j|}
    \sum_{l'=|J-j'|}^{|J+j'|}
    |\delta_{jj'}\delta_{ll'}
      \delta_{\nu \nu'}-S^{J}_{jj'll'\nu \nu'}|^2.
\end{multline}
The total energy $E$ is related to the kinetic energy of the
incoming particle according to $E=E_{\nu j}+\epsilon_{\nu j}$.

The rate coefficient for a given transition is obtained by
averaging the appropriate cross section over a Boltzmann
distribution of velocities of the projectile atom at a specific
temperature $T$:
\begin{equation}
k_{\nu j \rightarrow \nu' j'}(T)= G \int_{0}^{\infty}dE_{\nu
j}\sigma_{\nu j \rightarrow \nu' j'}(E_{\nu j})E_{\nu j}
e^{(-\beta E_{\nu j})},
\end{equation}
where the constant $G = \sqrt{\left(\frac{8}{\mu \pi
\beta}\right)}\beta^2$ and $\beta =(k_B T)^{-1}$ with $k_B$ being
Boltzmann's constant. The total quenching rate coefficient can
be calculated from
\begin{equation}
k_{\nu j}(T)= \sum_{\nu' j'} k_{\nu j \rightarrow \nu' j'}(T).
\end{equation}

\subsection{Potential energy surfaces of HeH$_2$}

Being one of the simplest triatomic molecular systems, the HeH$_2$
PES has been extensively studied theoretically over the last four
decades. Since each of He and H$_2$ has only two valence electrons
in closed shells, this trimer provides a fundamental test of the
quantum chemistry methods for calculating the van der Waals
interactions. The search for evidence of a bound HeH$_2$ halo
molecule is also a fascinating subject and its stability depends
sensitively on the PES. The PES has also been explored
experimentally using a variety of state-of-the-art experimental
techniques \cite{Toennies1,Toennies2}.

The historical development of the refinements of the HeH$_2$ PES,
which can be traced back to at least the early 1960s, has been
reviewed by Boothroyd, Martin, and Petersen \cite{BMP}. Here, we
outline some characteristics of the two most recently published
{\it ab initio} analytic HeH$_2$ potential energy surfaces, both
of which we have incorporated into our scattering theory computer
program. The first PES we used was published in 1994 by Muchnick
and Russek (MR) which was adopted in the scattering calculations
performed by Flower {\it et al.} \cite{Flower98}, Balakrishnan
{\it et al.} \cite{Bala98,Bala99} and Forrey {\it et al}.
\cite{Forrey98,Forrey99,Forrey01}. This {\it ab initio} HeH$_2$
PES overcame many of the deficiencies of earlier PESs which
adopted the rigid rotor approximation or considered only
equilibrium H$_2$ geometries. These earlier PESs also lacked fits,
either semiempirical or empirical, for the remainder of the
surface that was not constrained with {\it ab initio} data, which
severely limited their use for atom-diatom collision calculations.
However, MR generalized the HeH$_2$ PES based on the physics
underlying the principal interaction mechanisms responsible for
the PES, using 19 fitting parameters. This surface was fitted to a
combination of the {\it ab initio} energies of Meyer, Hariharan,
and Kutzelnigg \cite{Meyer} and of Russek and Garcia \cite{RG}. As
a result, the MR HeH$_2$ interaction was constructed to be
accurate in the van der Waals potential well and at the small-$R$
repulsive wall and to have physically reasonable behavior in
regions of the PES not constrained by {\it ab initio} data.

The second HeH$_2$ PES we employed is the most recent one
published by Boothroyd, Martin, and Petersen \cite{BMP} in 2003.
This PES was devised to represent accurately the van der Waals
well and the interaction region required for chemical reaction
dynamics. Consequently, a new set of  over 25,000 {\it ab initio}
points was calculated for HeH$_2$ geometries. Both the
ground-state and a few excited-state energies were computed, and
the conical intersection of the ground state with the first
excited state was mapped out approximately. A new analytic PES was
fitted to the {\it ab initio} data, yielding an improvement by
more than an order-of-magnitude in the fit in the interaction
region, compared to the MR HeH$_2$ PES. Unlike the MR PES, {\it ab
initio} points were used to constrain the fit for large H--H
separations.

Both the BMP and MR PESs are expressed as a function of distances
between the three atoms in the system. For the purpose of the
collision calculations, however, it is more convenient to expand
the interaction potential in terms of Legendre polynomials
$P_{\lambda}$ of order of $\lambda$ as in the MOLSCAT computer
program:
\begin{equation}
V_I(\vec{r},\vec{R}) = \sum^{\infty}_{\lambda}
v_{\lambda}(r,R)P_{\lambda}(cos \theta). \label{legpot}
\end{equation}
The reduced potential coupling matrix elements required for the
scattering calculations are obtained from
\begin{equation}
v^{\lambda}_{\nu j\rightarrow \nu' j'}(R)=\int_0^{\infty} dr
\chi^{*}_{\nu j}(r) v_{\lambda}(r,R) \chi_{\nu' j'}(r).
\label{vint}
\end{equation}
Neither BMP nor MR express their PES functions in the form of
eqn.(\ref{legpot}), but in terms of multi-body expansions with
physically-motivated functional forms.

\section{RESULTS AND DISCUSSION}\label{results}

We have carried out close-coupling calculations for collisions of
$^4$He with H$_2$ using the BMP and MR PESs. The total quenching
rate coefficient for H$_2$($v=1,j=0$) is shown in Fig.~1 for
temperatures in the range 10$^{-4}$ K and 300 K. The rate coefficients
attain finite values for temperatures lower than 1 mK in
accordance with Wigner's law.  Unexpectedly, we find that the
total quenching rate coefficient computed with the BMP PES is as
much as three orders of magnitude larger than that calculated with
the PES of MR. Only the results from the MR PES agree with the
experimental data of Audibert {\it et al.} \cite{Audibert} Good
agreement with the Audibert {\it et al.} results were previously
obtained by Balakrishnan {\it et al.} \cite{Bala98} who also used
the MR PES. This significant discrepancy suggests that the PES of
BMP may contain some unphysical behavior in either the newly
computed {\it ab initio} data or the adopted fit functions.
The different slopes of the rate coefficients obtained using
the BMP and MR surfaces indicate
that a much larger fraction of the available energy is taken up as rotation  in
collisions on the BMP surface.  To
identify the origins of this discrepancy in the BMP surface, we
have examined the constituent channels of the quenching rate
coefficients.

Fig.~2 displays the inelastic rovibrational state-to-state,
non-thermal rate coefficients as a function of collision energy.
The non-thermal rate coefficient is defined here as cross section
$\sigma_{1,0\rightarrow0,j'}$ $\times$ collision velocity
\textsl{v} in the center of mass frame. The rate coefficients
computed using the BMP PES (see Fig.~2a) are seen to increase
monotonically with $j'$ over the entire energy range. This
behavior is in sharp contrast to the state-to-state rate
coefficients computed with the MR surface shown in Fig.~2b which
display no obvious ordering with $j'$. The discrepancy between the
total quenching rate coefficients shown in Fig.~1 is evidently due
to the $j'$=8 channel with the BMP surface the rate coefficient of
which is more than three orders of magnitude larger than any of
the state-to-state MR rate coefficients given in Fig.~2b. An
appreciable contribution also comes from the $j'$=6 channel with
the BMP result being a factor of 10 larger than that obtained with
the MR PES.

By calculating the $\lambda$-dependent potential couplings on
eqn.(\ref{legpot}), we may gain some insight into the origin of
the discrepancies. Fig.~3 shows diagonal and off-diagonal reduced
potential coupling matrix elements, defined in eqn. (\ref{vint}).
Only the even terms contribute here since the diatom is a
homonuclear molecule. To analyze the present results, which depend
primarily on the intermediate and long range behavior of the PES,
we examine the first five terms ($\lambda$ = 0, 2, 4, 6, 8) in the
expansion of $V_I$($\vec{r},\vec{R}$). Figs.~3a and 3b show a
comparison between the BMP and MR diagonal matrix elements for
($\nu=1, j=0$). The first radial coefficient $v^0(R)$ represents
the spherical component of the interaction, while the higher terms
define the anisotropic behavior of the PES. Comparing $v^0(R)$
obtained from the BMP and MR PES, we find that the two yield
essentially identical results with the repulsive wall of the
($\nu=1, j=0$) channel for both cases occurring at distances just
less than $R$ = 6 a.u. While there are evidently significant
differences for the higher $v^{\lambda}$($R$) terms for $R < 5$
a.u., these differences appear inside the repulsive barrier (the
$v^0$($R$) term), a region which will not be accessible in the
low-energy collisions considered here, and therefore will have
negligible impact on the cross sections.

In Figs.~3c and 3d, the off-diagonal coupling matrix elements for
the dominant state-to-state transitions obtained with the BMP
surface are presented. Fig.~3c reveals the striking difference
between the BMP and MR off-diagonal coupling matrix element for
$\lambda = 8$. Comparing $\lambda=0$ of Fig.~3a with Fig.~3c, we
see that the BMP potential provides a significant off-diagonal
$\lambda=8$ coupling for distances outside the potential barrier
which are accessible during the collision. Conversely, the MR
coupling is shifted to smaller $R$ so that the magnitude of the
coupling is significantly smaller in the interaction region.
Therefore, the discrepancy  in the (1,0)$\rightarrow$(0,8)
quenching rate coefficient, as well as the total, can be
attributed to the enhancement by the BMP surface of the
$\lambda=8$ off-diagonal coupling. Fig.~3d makes a similar
comparison of off-diagonal couplings for $\lambda=6$. The BMP
coupling is again larger, but the difference is not as dramatic as
for the $\lambda=8$ case, consistent with the smaller differences
between the (1,0)$\rightarrow$(0,6) transition rate coefficients
shown in Fig.~2.

The importance of including higher-order terms in the  Legendre
polynomial expansion of the interaction potential in accurately
determining vibrational quenching rate coefficients at low
temperatures has been the topic of a number of recent
investigations \cite{krems02a,krems02b,bala03,uudus04}. Krems
\cite{krems02a} showed that for the vibrational relaxation of
CO($\nu=1$) by collisions with He atoms terms as high as
$\lambda=30$ were required in eq.(\ref{legpot}) to obtain
converged cross sections. In subsequent calculations, Krems {\it
et al.} \cite{krems02b} reported similar behavior for the
vibrational quenching of HF($\nu=1,j=0$) by collisions with Ar
atoms which  preferentially populate high lying rotational levels
in the $\nu=0$ vibrational level. In a more recent study Uudus
{\it et al.} \cite{uudus04} reported significant differences in
the low temperature vibrational relaxation  rates of
H$_2(\nu=1,j=0)$ by collisions with Ar atoms computed using two
different interaction potentials. They found that
(1,0)$\rightarrow$(0,8) transition dominates the quenching when
the Ar-H$_2$ potential of Bissonnette {\it et al.} \cite{biss96}
is used while no such preference is observed when the interaction
potential of Schwenke {\it et al.} \cite{schwenke93} is employed.
As in the present study, the differences were attributed to
enhanced off-diagonal coupling arising from the $\lambda=8$ term
of the potential surface of Bissonnette {\it et al.} \cite{biss96}

The dominance of the (1,0)$\rightarrow$(0,8) transition found
using the BMP PES is a consequence of two characteristics: (i) the
small energy gap between the initial and final states (the energy
gap is only 111 cm$^{-1}$) and (ii) the relatively large potential
coupling shown in Fig.~3c. In previous studies using the MR
potential (i) was satisfied, but not (ii) so that the importance
of this channel was not seen. Using the BMP potential, we can
demonstrate that both characteristics are required to obtain a
large transition rate by artificially enlarging the energy gap by
increasing the target molecular reduced mass. In Figs.~4, the
zero-temperature rate coefficient for the
(1,0)$\rightarrow$(0,$j'$) transitions obtained with the BMP
potential are displayed as a function of target molecule reduced
mass. The nearly exponential decrease in the
(1,0)$\rightarrow$(0,8) transition, and the smaller decreases in
the other channels, is consistent with the above argument. While
there is no change in the BMP potential coupling, (i) is reduced
with the increasing energy gap (increasing reduced mass) until the
(1,0)$\rightarrow$(0,8) channel loses its dominance for
$\frac{m}{m_{H_2}}>2$. For $\frac{m}{m_{H_2}}\sim 2.5$ (where 2.4
corresponds to DT), the $j'$=10 channel becomes exoergic and its
dominance is likely to be due to an enhanced BMP off-diagonal
coupling for $\lambda=10$.

To ascertain whether the above discrepancy will manifest itself
for other inelastic transitions, we also calculated the
non-thermal rate coefficients for quenching of the initial
($\nu=2, j=0$) state to individual rotational levels of the
$\nu=1$ vibrational state using both the BMP and MR potential
surfaces shown in Fig.~5. The dependence on $j'$ of the
transitions is identical to that seen for the ($\nu=1, j=0$)
quenching rate coefficients for the two potentials. Again, the
major contribution comes from the $j'=8$ channel when the BMP
potential is used.

For the higher vibrational states, we illustrate in Figs.~6a and
6b the variation of the total quenching rate coefficients with
respect to the vibrational quantum number $\nu$ for $j=0$ and
$j=1$ at zero temperature. We found that the $j'=8$ component is
unmistakably dominant for $\nu < 4$ when the BMP PES is used.
However, when $\nu > 4$, the $j'=8$ channel is closed and the
total quenching rate is dominated by the $j'=6$ contribution. The
peak observed at $\nu = 10$ is due to a zero-energy resonance
which is a consequence of a quasi-bound state of the HeH$_2$
complex (see below).

For $j$=1, shown in Fig.~6b, there is no $j'$=8 contribution since
transitions to this state are forbidden and there are no $j'$=9
contributions because the channel is closed for all $\nu$. The
major contribution to the total quenching rate coefficients comes
from the $j'=7$ component ($\lambda=6$) instead. We found that the
$j'=7$ contribution for $j=1$ continues to be strong up to high
energies; the zero-energy resonance is also present in the BMP PES
for $\nu = 11, j=1$.

In Figs.~7a, 7b, 8a, and 8b we illustrate the energy-dependent
elastic cross sections for $\nu$=0, 1, 2, and 10, respectively for
both the MR and BMP potentials and with $j=0$. The differences
seen for the first three are likely due to improvements in the
long-range portion of the BMP potential compared to that of the MR
surface where anisotropy effects do not play a role. The two
potentials give good agreement for elastic scattering at high
collision energies.

For the ($\nu=10, j=0$) elastic cross section, the results based
on the two PESs give reasonable agreement at high collision
energies. However, as the collision energy decreases below $E
\sim$ 0.1 cm$^{-1}$, the two curves start to deviate considerably.
Both a significant enhancement of the elastic cross section with
decreasing collision energy and a shift in the onset of the Wigner
threshold behavior in the ultracold region can be attributed to
the presence of a zero-energy resonance in the BMP PES-based
calculation. This situation, where one PES-based calculation
predicts a zero-energy resonance and the other does not, was seen
before in the case of Ar+H$_2$ collisions \cite{Flasher}.
Because the van der Waals potential well supports a weak
 quasi-bound state
for each excited level of the diatom \cite{Forrey98}, a small change in the PES can
easily move the quasi-bound state into coincidence with the
corresponding diatomic energy level.
Therefore, the existence or absence of such a resonance is a very
sensitive test of the details of the PES.

Returning to inelastic collisions, other differences that are
manifested may be seen by considering the $\nu=10, j=0$ state
since a peak in the $\nu$-distribution for the zero-energy
quenching rate coefficient is evident in Fig.~6a. Specifically,
Figs.~9a and 9b show the energy dependence of the rate
coefficients for transitions from the initial ($\nu=10, j=0$)
level into ($\nu=9, j'$) for the BMP and MR potentials,
respectively. Fig.~9a shows that for collision energies less than
$\sim$0.1 cm$^{-1}$, the rate coefficients increase with
decreasing energy. This is due to the presence of the zero-energy
resonance. However, the non-thermal rate coefficients computed
using the MR potential shown in Fig.~9b do not display this
behavior -- the zero-energy resonance simply does not exist for
the MR potential. Therefore the ultracold quenching cross sections
are very different for the MR and BMP PESs. Nevertheless, both
results are in good agreement at $E \sim$ 100 cm$^{-1}$, which is
expected for high collision energies, since the influence of the
zero-energy resonance gradually decreases with energy.

This differs from the strong $j'=8$ contribution from the BMP PES
for $\nu < 4$ which persists even for higher energies (as
demonstrated in Fig.~2). Note further that the $j'$-orderings of
the ($\nu=10, j=0$) state-to-state rate coefficients are not the
same as observed for the ($\nu=1, j=0$) and ($\nu=2, j=0$) cases.

From these results, we find that the largest discrepancies
occurred for vibrational transitions where large changes in
angular momentum are allowed. Such transitions probe the
anisotropy of the potential to high order. The differences noted
for the elastic processes, which are determined by the spherical
($\lambda=0$) term of the potential, are likely due to
improvements made in the BMP PES.

Figs.~10a and 10b display the (0,2)$\rightarrow$(0,0) and
(1,2)$\rightarrow$(1,0) deexcitation cross sections for pure
rotational transitions. The results from both PESs are very
similar. The BMP result always being 10--30\% smaller. The
discrepancy being $\sim$10\% in the high energy region. These
transitions are not strongly affected by the higher-order
anisotropy of the potential (i.e., $\lambda \ge 2$), and the BMP
results are likely to be improvements. This is further illustrated
in Fig.~11 where the zero-temperature total quenching rate
coefficients for $\nu=1$ are given as a function of $j$. For $j$
$>$ 1, pure rotational quenching dominates and the two PESs give
results which agree to within about a factor of two. The
significant drop in the $j=22$ and 23 quenching rate coefficients
was explored previously by Forrey \cite{Forrey01}.

\section{SUMMARY AND CONCLUSIONS}\label{summary}

In summary, we have carefully and critically compared the two most
recently published {\it ab initio} interaction potential energy
surfaces (PESs) for HeH$_2$ by performing elastic and inelastic
scattering calculations for collisions of $^4$He with H$_2$. The
calculations were performed using a non-reactive
quantum-mechanical close-coupling method and were carried out for
collision energies ranging from the ultracold (10$^{-6}$
cm$^{-1}$) to the thermal (100 cm$^{-1}$) regime. Using the
Muchnick and Russek \cite{MR} PES, we have reproduced the
inelastic cross sections $\sigma_{10 \rightarrow 0j'}(E)$ and the
total quenching rate coefficients of Balakrishnan {\it et al.}
\cite{Bala98}. However, the inelastic cross sections $\sigma_{10
\rightarrow 0j'}(E)$ obtained from the Boothroyd {\it et al.}
\cite{BMP} PES are significantly different. The corresponding
total quenching rate coefficients turn out to be a thousand times
larger than those obtained with the MR surface that agree well
with the measurements of Audibert {\it et al.} \cite{Audibert}. We
attribute this discrepancy to an enhancement, possibly unphysical,
in a high-order anisotropy component of the BMP potential which is
manifested primarily in the ($\nu=1, j=0$) to ($\nu'=0, j'=8$)
transition. The effect, which is a combination of a sizeable
overlap of wavefunctions for transitions between states with a
small energy gap and a large anisotropic potential coupling, is
observed for the total quenching cross section from other ($\nu,
j=0$) states until the $j'=8$ channel becomes closed. The effect
is suppressed when either the energy gap is increased (e.g.
 by increasing the molecular target reduced mass) or when pure
rotational transitions dominate the quenching. These results lead
us to postulate, primarily based on the discrepancy with
experiment, that the high-order anisotropy components of the BMP
potential surface are not accurately determined and make the PES
unsuitable for studies of vibrational transitions.

We also find zero-energy resonances in the BMP potential which
result in significant enhancements to both the elastic and
inelastic cross sections. There are, however, no experimental data
which could help to determine the reality of the zero-energy
resonances. While the ($\nu=10, j=0$) elastic cross sections for
both PESs agree well at high energies, the results are
considerably different at ultracold energies because of the
manifestation of the zero-energy resonance for the BMP surface,
causing a shift in the onset of the Wigner threshold behavior in
this energy regime.

Finally, for pure rotational transitions, the cross sections
obtained with the two PESs agree within
a factor of two. This difference may actually signify
improvements in the BMP surface for the less anisotropic
components of the potential. In conclusion, the BMP surface will
need to be reevaluated before it can be adopted in large-scale
scattering calculations. Further experiments are needed for
inelastic and elastic processes to aid in resolving these issues
as well as for benchmarking the scattering calculations.

\section{ACKNOWLEDGMENTS}
T.G.L. acknowledges support from the University of Kentucky with
additional support from ORNL and the University of Georgia; and
general discussions with Dr. Predrag Kristic. The work of C.R.,
R.M., T.K.C., and R.C.F. was supported by NSF grant PHY-0244066.
N.B. acknowledges support from NSF grant PHY-0245019. P.C.S.
acknowledges support from NSF grant AST-0087172 and helpful
discussions with Stephen Lepp. A.D. is supported by the Chemical
Sciences, Geosciences and Biosciences Division of the Office of
Basic Energy Sciences, US Department of Energy.

\newpage

\begin{figure}[h]
\includegraphics[width=\figwidth]{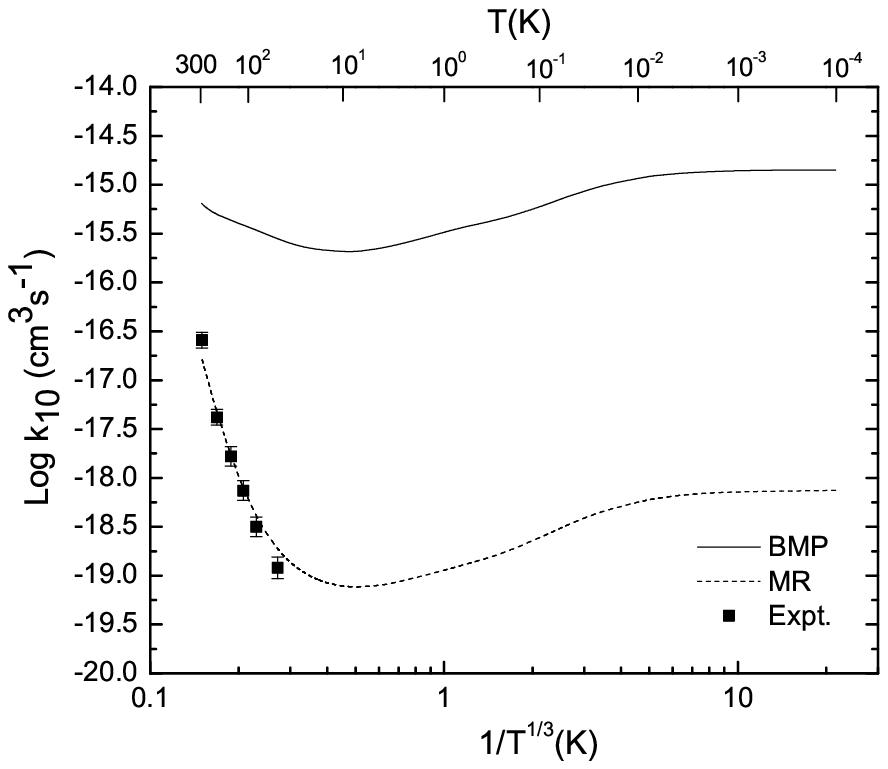}
\caption{Total quenching rate coefficient for ($v=1,j=0$) as a
function of temperature. The solid and dashed curves are results
obtained using the BMP and MR potentials, respectively. The solid
squares with error bars are the experimental results of Audibert
{\it et al.} \cite{Audibert}. The MR curve agrees very well with
the experimental data whereas the BMP curve does not.}
\end{figure}

\begin{figure}[h]
\includegraphics[width=\figwidth]{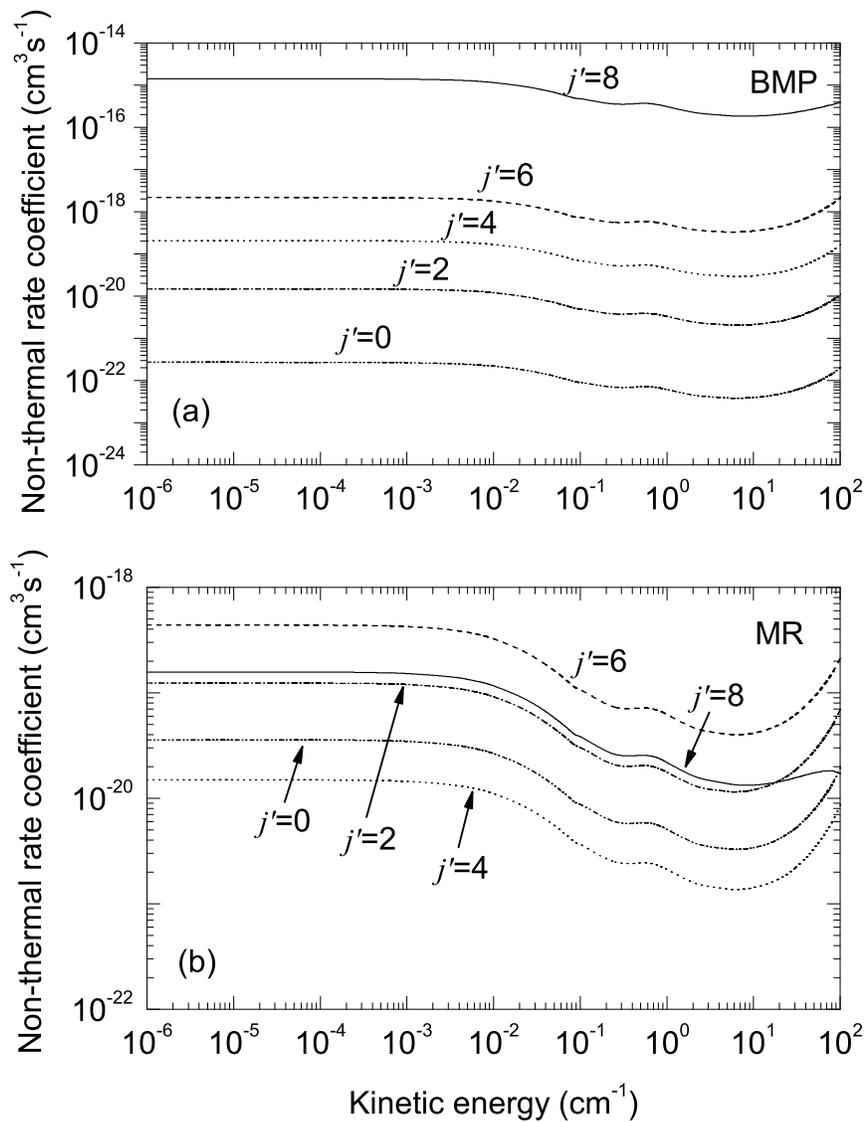}
\caption{Inelastic cross section $\sigma_{1,0\rightarrow 0,j'}$
times collision velocity (non-thermal rate coefficient) as a
function of collision energy. Computed using the BMP (a) and MR
(b) PES. The $j'=8$ contribution strongly dominates the BMP
results over the entire energy range shown and is responsible for
the large discrepancy with experiment (see Fig. 1).}
\end{figure}

\begin{figure}[h]
\includegraphics[width=\figwidth]{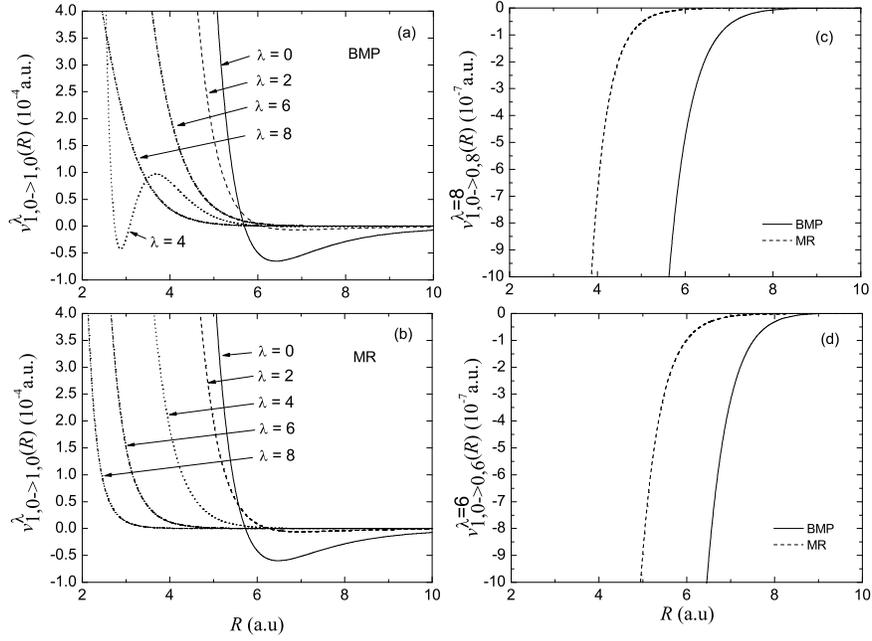}
\caption{(a) Diagonal matrix element for ($\nu=1, j=0$) using the
BMP potential. The curves correspond to the $\lambda=0, 2, 4, 6,
8$ terms of a Legendre expansion of the potential energy surface
(see eqn.(\ref{vint})). (b) Same as (a) but using the MR
potential. (c) Off-diagonal coupling matrix element for
$\lambda=8$. (d) Off-diagonal coupling matrix element for
$\lambda=6$.}
\end{figure}

\begin{figure}[h]
\includegraphics[width=\figwidth]{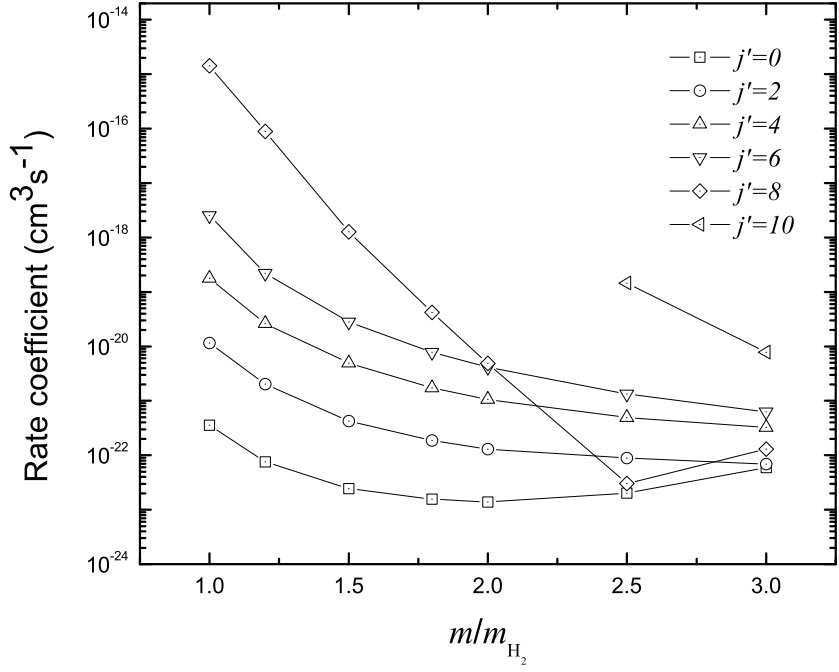}
\caption{Zero-temperature rate coefficients for ($\nu=1, j=0$) as
a function of diatom reduced mass, $m$, using the BMP potential.
The nearly exponential behavior of the $j'=8$ contribution is
consistent with exponential energy gap behavior seen previously
\cite{Forrey99}. The inelastic results for $m$ corresponding to DT
and T$_2$ are not as sensitive to the $j'=8$ contribution.}
\end{figure}

\begin{figure}[h]
\includegraphics[width=\figwidth]{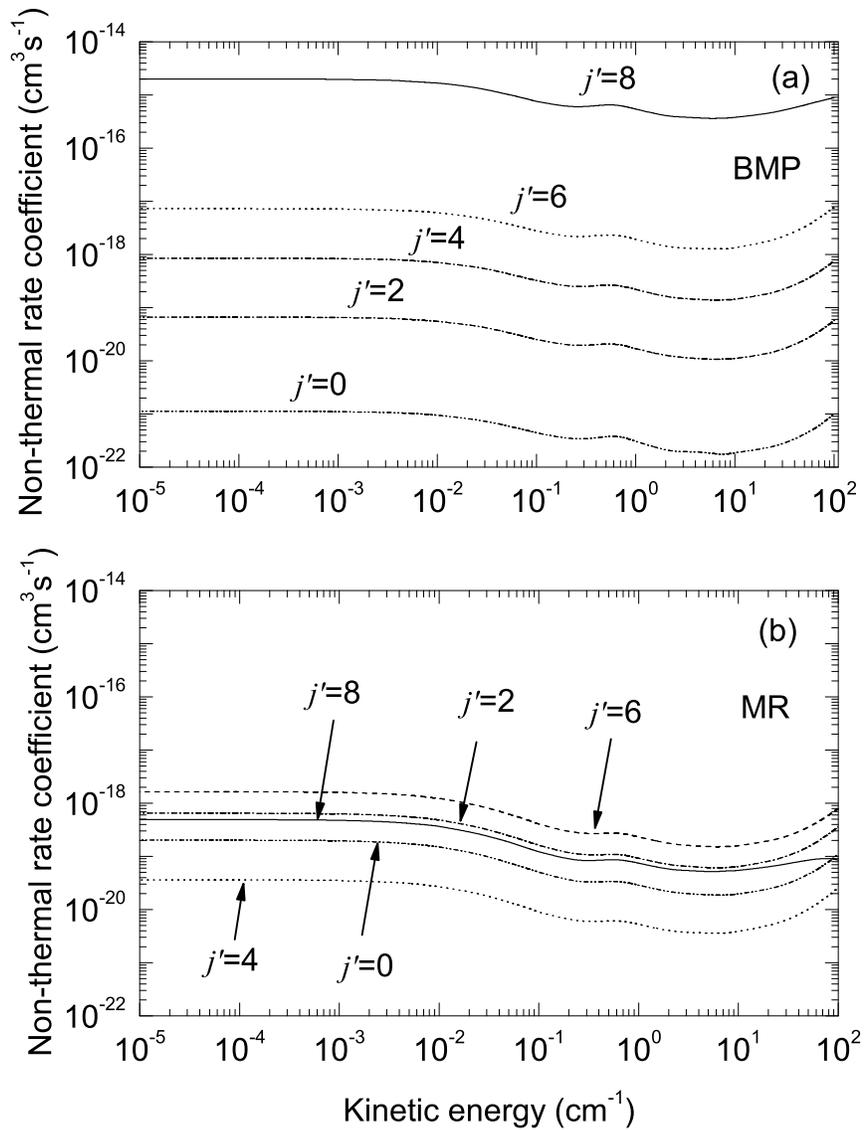}
\caption{Inelastic cross section $\sigma_{2,0\rightarrow 1,j'}$
times collision velocity (non-thermal rate coefficient) as a
function of collision energy computed with the BMP (a) and the MR
(b) potential. The $j'=8$ contribution again strongly dominates
the BMP results and produces a significant disagreement with the
MR results.}
\end{figure}

\begin{figure}[h]
\includegraphics[width=\figwidth]{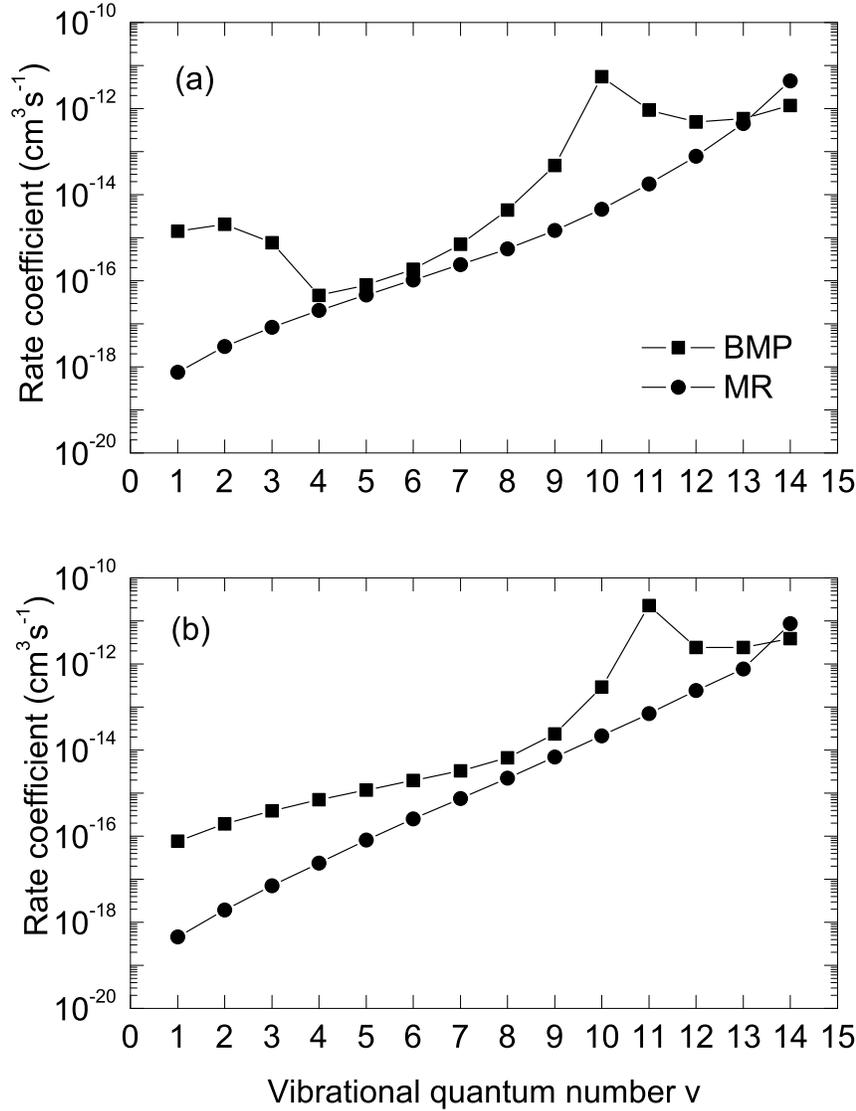}
\caption{(a) Zero-temperature total quenching rate coefficients
for $j=0$ as a function of $\nu$. The $j'=8$ contribution is
strongly dominant for $\nu <4$ when using the BMP potential. For
$\nu >4$ the $j'=8$ channel is closed and the total quenching rate
is dominated by the $j'=6$ contribution. (b) Same as (a) but for
$j$ = 1 case. The $j'=7$ contribution is dominant for low $\nu$
when using the BMP potential. The peaks in the BMP curves at
$\nu=10, j=0$ and $\nu=11, j=1$ are due to zero-energy resonances.
The influence of the zero-energy resonances disappears as the
collision energy is increased. The large differences between the
two potentials at low $\nu$, however, persist for all energies.}
\end{figure}

\begin{figure}[h]
\includegraphics[width=\figwidth]{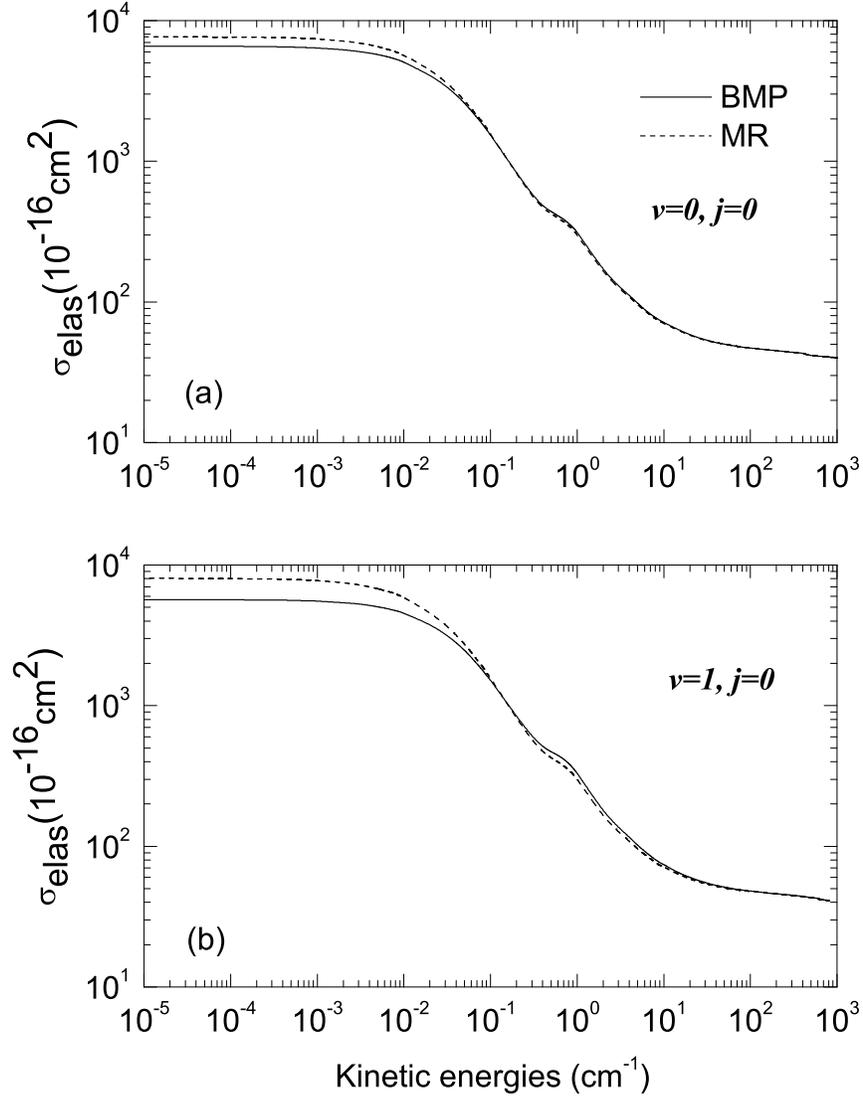}
\caption{Elastic cross sections $\sigma_{0,0}$ (a) and
$\sigma_{1,0}$ (b) as a function of collision energy. The two
potentials give good agreement for elastic scattering at high
collision energies. The difference at ultracold energies is within
a factor of 2 which is typical for most of the $(v,j)$ levels of
this system.}
\end{figure}

\begin{figure}[h]
\includegraphics[width=\figwidth]{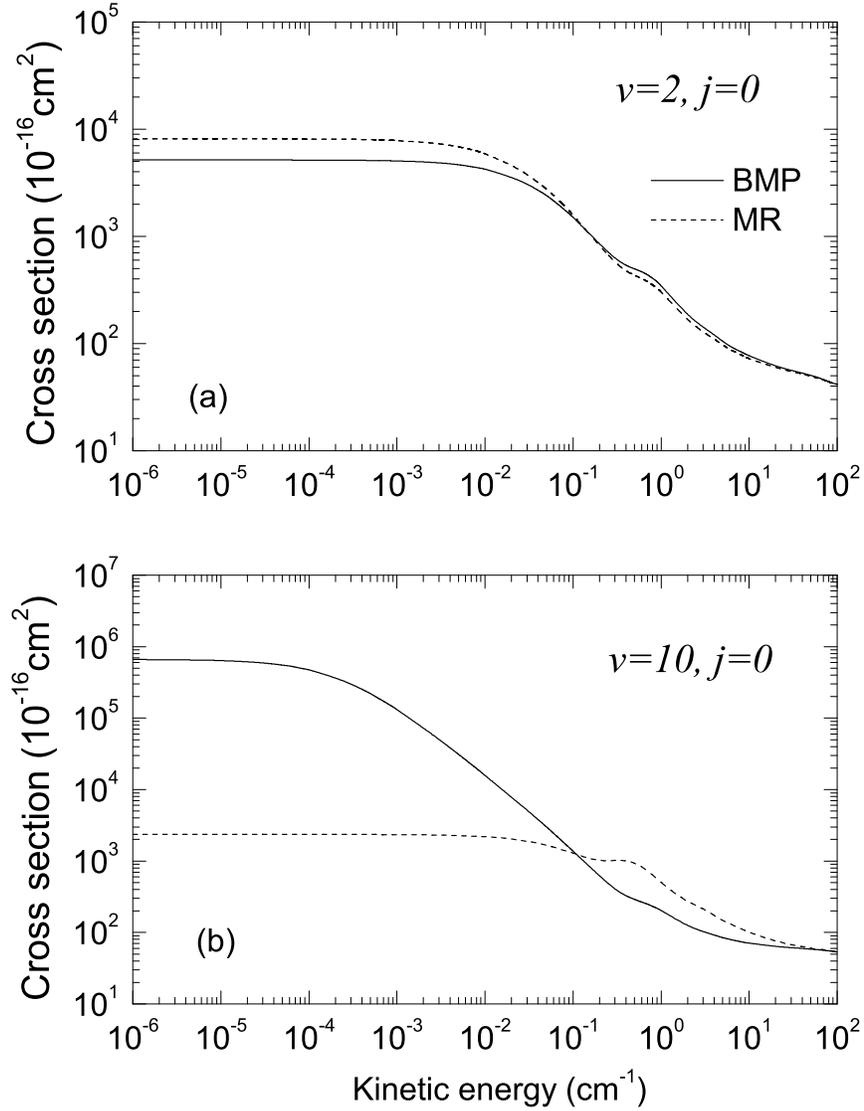}
\caption{Elastic cross sections $\sigma_{2,0}$ (a) and
$\sigma_{10,0}$ (b) as a function of collision energy. The two
potentials again give good agreement for elastic scattering at
high collision energies. The presence of a zero-energy resonance
for $(\nu=10, j=0)$ causes a significant increase in the BMP
result at low energies and a shift in the onset of the Wigner
threshold behavior at ultracold energies.}
\end{figure}

\begin{figure}[h]
\includegraphics[width=\figwidth]{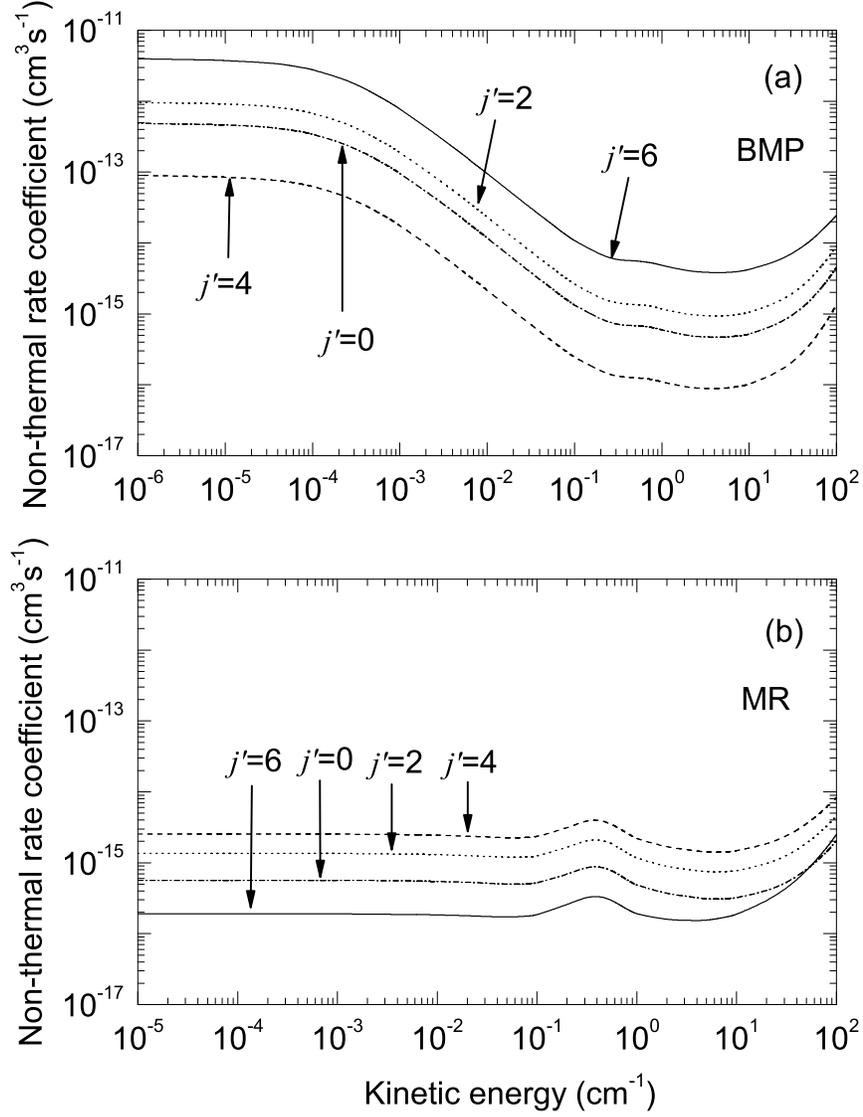}
\caption{(a) Inelastic cross section $\sigma_{10,0\rightarrow
9,j'}$ times collision velocity (non-thermal rate coefficient) as
a function of collision energy computed with the BMP potential.
The increase in the cross sections with decreasing energy is due
to the presence of a zero-energy resonance. (b) Same as (a) but
with the MR potential. There is no zero-energy resonance for the
MR potential, so the ultracold results are very different than in
Fig. 9a.}
\end{figure}

\begin{figure}[h]
\includegraphics[width=\figwidth]{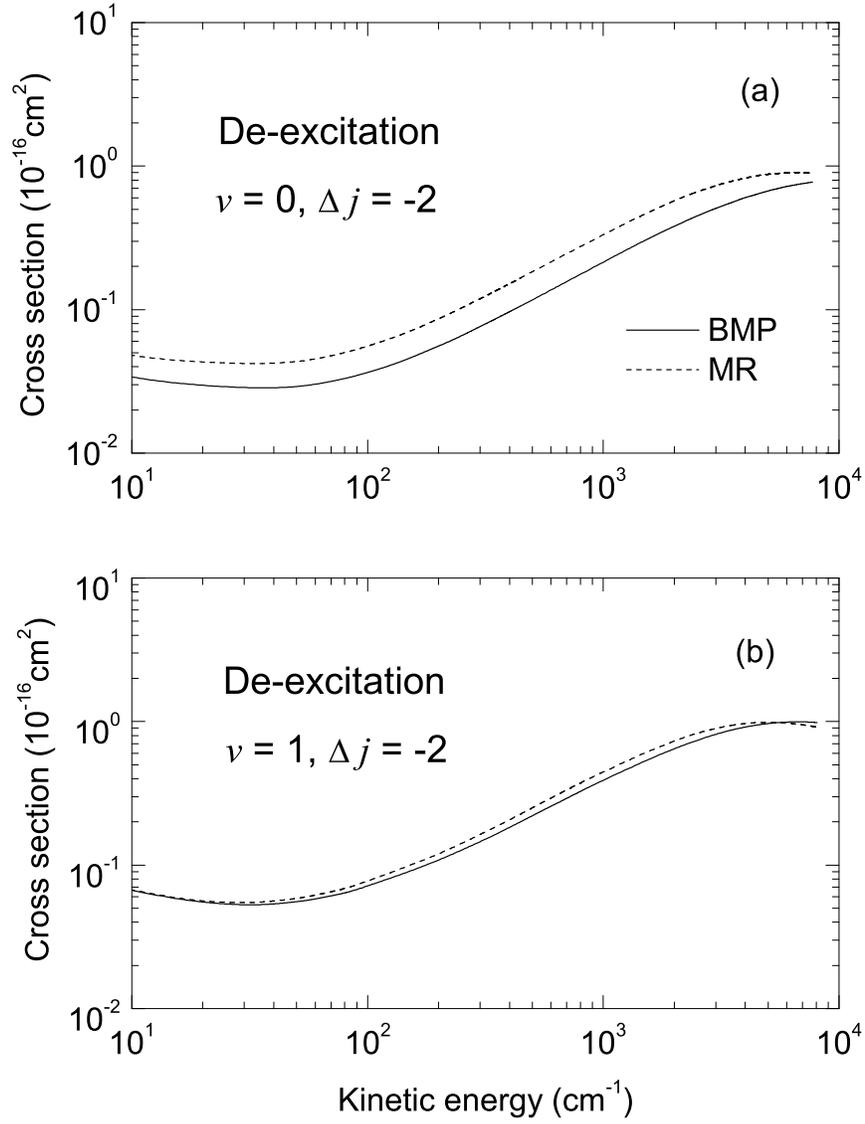}
\caption{Comparison of BMP and MR based de-excitation cross
section for the $\triangle j=-2$ rotational transition from the
$j=2$ initial rotational level of H$_2$ in vibrational levels of
$\nu$ = 0 and 1 as a function of collision energy.}
\end{figure}

\begin{figure}[h]
\includegraphics[width=\figwidth]{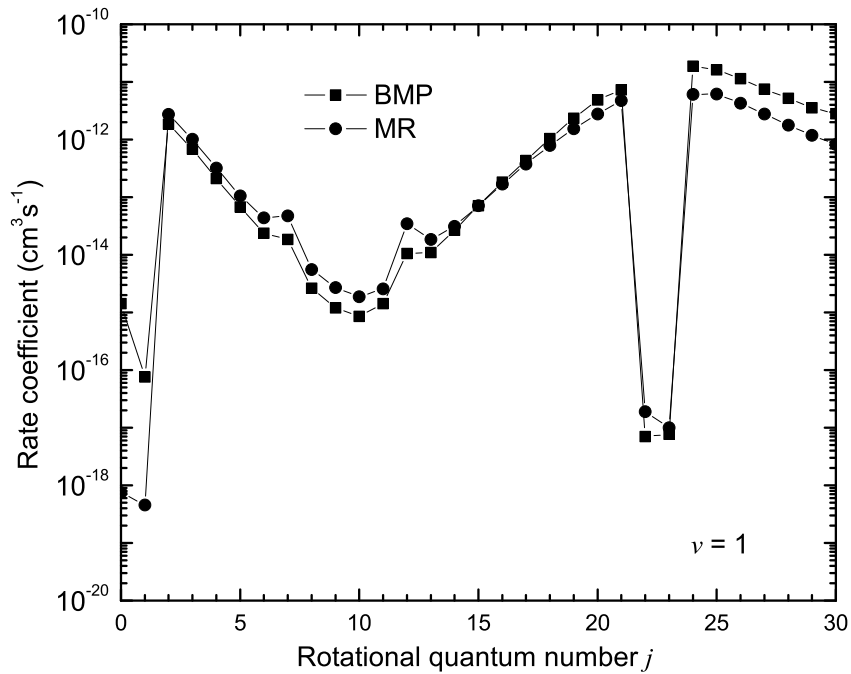}
\caption{Zero-temperature total quenching rate coefficients for
$\nu = 1$ as a function of rotational quantum number $j$. For
$j>1$, pure rotational quenching is allowed and the two potentials
give better agreement. The $j$-dependence of the rate coefficients
has been explored in detail by Forrey \cite{Forrey01}.}
\end{figure}

\end{document}